\documentclass[preprint]{aastex}
\usepackage{dcolumn}
\begin{document}
\title{Gravitational Redshift in Kerr - Newman Geometry}

\shorttitle{Gravitational Redshift in Kerr - Newman Geometry}
\shortauthors{A. K. Dubey and A. K. Sen}

\author{Anuj Kumar Dubey\altaffilmark{}}
\affil{Department of Physics, Assam University, Silchar, 788011, Assam, India}
\email{danuj67@gmail.com}
\and
\author{A. K. Sen \altaffilmark{}}
\affil{Department of Physics, Assam University, Silchar, 788011, Assam, India}
\email{asokesen@yahoo.com} 


\begin{abstract}
It is well known fact that gravitational mass can alter the space time structure and gravitational redshift  is  one of its  examples. Static electric or magnetic charge can also alter the space time structure,  similar to gravitational mass,  giving rise to its effect on redshift. This can also be considered as electro and magneto static redshift.  Gravitational redshift has been reported by most of the authors without consideration of static electric and / or magnetic charges present in the rotating body. In the present paper, we considered the three parameters: mass, rotation parameter and charge to discuss their  combined effect on redshift,  for a charged rotating body by using Kerr - Newman metric.  It has been found that,  the presence  of  electrostatic and magnetostatic charge increases the value of so-called gravitational redshift.  Calculations have been also done here to determine   the effect of electrostatic and magnetostatic charges on the amount of redshift of a light ray emitted at various latitudes from a charged rotating body. The variation of gravitational redshift from equatorial to  non- equatorial region has been calculated, for a given set of values of electrostatic and magnetostatic charges.
\end{abstract}
\section{Introduction}
\label{intro}
 Gravitational redshift has been reported by most of the authors without consideration of rotation of a body, as detailed in our previous work Dubey and Sen (2014). Payandeh and Fathi (2013) had obtained the gravitational redshift for a static spherically symmetric electrically charged object in Isotropic Reissner - Nordstrom Geometry. Dubey and Sen (2014) had obtained the expression for gravitational redshift from rotating body in Kerr Field. They also showed the rotation and the latitude dependence of gravitational redshift from a rotating body (such as pulsars). The expression of Gravitational Redshift Factor ($\Re$) from rotating body in Kerr Field was given as (Eqn. 69, Dubey and Sen (2014)):
\begin{equation}
\Re(\phi ,\theta)= \sqrt{{g_{tt}+ g_{\phi\phi}(\frac{d\phi}{c dt})^{2}} +2 g_{t \phi}(\frac{d\phi}{c dt})}
\end{equation}
  Apsel (1978, 1979) had discussed, that the motion of a particle in a combination of gravitational and electro and magneto static field can be determined from a variation principle of the form $\delta\int d\tau$ =0. The form of the physical time is determined from an examination of the Maxwell - Einstein action function. The field and motion equations are actually identical to those of Maxwell - Einstein theory. The theory predicted that even in a field free region of space, electro and magneto static potentials can alter the phase of wave function and the life time of charged particle.\\
 Gravitational redshift has been reported by most of the authors without consideration of static electric and / or magnetic charge present within the rotating body. With this background, the present paper is a continuation of our previous work Dubey and Sen (2014), to study the influence of electro and magneto static charges of a rotating gravitating body on the redshift.\\
The present paper is organized as follows. In Section - 2, we  derive the expression for radial component of electro and magneto static field from charged rotating body by using the associated potential of Kerr-Newman Field. In Section -3, we derive the expression for combined gravitational and electro-magneto-static redshift from rotating body. Finally, some conclusions are made in Section -4.
\section{\label{sec: electro-magneto static charge} Radial Component of Electro and Magneto Static Field from Charged Rotating Body}
When rotation is taken into consideration spherical symmetry is lost and off - diagonal terms appear in the metric and the most useful form of the solution of Kerr family is given in terms of t, r, $\theta$ and $\phi$, where t, and r are Boyer - Lindquist coordinates running from - $\infty$ to + $\infty$, $\theta$  and $\phi$, are ordinary spherical coordinates in which  $\phi$ is periodic with period of 2 $\pi $ and $\theta$ runs from 0 to $\pi$.
 Covariant form of metric tensor for Kerr family (Kerr (1963), Newman et al. (1965)) in terms of Boyer-Lindquist coordinates with signature (+,-,-,-) is expressed as:
\begin{equation}
ds^{2} = g_{tt}c^{2}dt^{2}+ g_{rr} dr^{2} + g_{\theta\theta} d\theta^{2} +g_{\phi\phi} d\phi^{2} + 2 g_{t \phi} c dt d\phi
\end{equation}
where $g_{ij}$'s are non-zero components of Kerr family. \\
If we consider the three parameters: mass (M), rotation parameter (a) and charge (electric (Q) and / or magnetic (P)), then it is easy to include charge in the Non-zero components of $g_{ij}$ of Kerr metric, simply by replacing $(r_{g} r)$ with $(r_{g} r-Q^{2}-P^{2})$.
 \\ Non-zero components of $g_{ij}$ of Kerr-Newman metric are given as follows (page 261-262 of Carroll (2004)):
\begin{equation}
 g_{tt} = (1-\frac{r_{g} r-Q^{2}-P^{2}}{\rho^{2}})
\end{equation}
\begin{equation}
g_{rr}=-\frac{\rho^{2}}{\Delta}
\end{equation}
 \begin{equation}
 g_{\theta\theta}=-{\rho^{2}}
 \end{equation}
 \begin{equation}
g_{\phi\phi}= -[r^{2}+a^{2}+\frac{(r_{g}r-Q^{2}-P^{2}) a^{2}sin^{2}\theta}{\rho^{2}}]sin^{2}\theta
 \end{equation}
\begin{equation}
g_{t\phi}=\frac{a sin^{2}\theta (r_{g}r-Q^{2}-P^{2})}{\rho^{2}}
\end{equation}
with
\begin{equation}
\rho^{2} = r^{2}+ a^{2}cos^{2}\theta
\end{equation}
and
\begin{equation}
 \Delta =r^{2}+ a^{2}-r_{g}r +Q^{2}+P^{2}
\end{equation}
where $r_g=(2GM/c^2)$ is the Schwarzschild radius. Q and P are electric and magnetic charges respectively and a $(= \frac{J}{Mc})$ is rotation parameter of the source. If we replace $(r_{g} r-Q^{2}-P^{2})$ by $(r_{g} r-Q^{2})$ and further if we put rotation parameter of the source (a) equal to zero, then it reduces to Reissner - Nordstrom metric. Also  if we replace $(r_{g} r-Q^{2}-P^{2})$ by $(r_{g} r)$ then the Kerr-Newman metric reduces to Kerr metric and further if we put rotation parameter of the source (a) equal to zero then it reduces to Schwarzschild metric.\\
The associated potential of the Kerr-Newman metric are expressed as (page 262 of Carroll (2004)):
\begin{equation}
A_{t} = \frac{Q r - P a cos\theta}{r^{2}+ a^{2} cos^{2}\theta}
\end{equation}
\begin{equation}
A_{r} = 0
\end{equation}
\begin{equation}
A_{\theta} = 0
\end{equation}
\begin{equation}
A_{\phi} = \frac{-Q a r sin^{2}\theta + P (r^{2}+ a^{2}) cos\theta}{r^{2}+ a^{2} cos^{2}\theta}
\end{equation}
Electromagnetic field strength tensor $(F_{ij})$ can be expressed in terms of potential as (page 65 (Eqn. 23.3) of Landau and Lifshitz (2008)):
\begin{equation}
F_{ij}=\frac{\partial A_{j}}{\partial x^{i}}-\frac{\partial A_{i}}{\partial x^{j}}
\end{equation}
Radial component of Electric Field $(E^{r})$ can be related to electromagnetic field strength tensor $(F_{ij})$ by the expression (page 254 of Carroll (2004)):
\begin{equation}
E^{r}=F_{rt}=-F_{tr}
\end{equation}
Using equations (10-14) we can express the radial component of electric field $E^{r}$ as:
\begin{equation}
 E^{r}=F_{rt}=\frac{\partial A_{t}}{\partial{r}}-\frac{\partial A_{r}}{\partial{t}}=\frac{\partial A_{t}}{\partial{r}}
\end{equation}
After simplification the above equation can be written as:
\begin{equation}
 E^{r}=F_{rt}=\frac{Q}{r^{2}+ a^{2}cos^{2}\theta}-\frac{2r(Qr-Pa cos\theta)}{(r^{2}+ a^{2}cos^{2}\theta)^{2}}
\end{equation}
Radial component of Magnetic field $(B^{r})$ can be related to electromagnetic field strength tensor $(F_{ij})$ by the expression (page 254 of Carroll (2004)):
\begin{equation}
B^{r}=\varepsilon^{trij} F_{ij}
\end{equation}
Levi-Civita Tensor ($\varepsilon^{\rho\sigma ij}$) and Levi-Civita Symbol ($\tilde{\varepsilon}^{\rho\sigma ij}$) in four - dimension are related by the expression (page 24 and 83 of Carroll (2004)):
\begin{equation}
\varepsilon^{\rho\sigma ij} = \frac{\tilde{\varepsilon}^{\rho\sigma ij}}{\sqrt{-g}}
\end{equation}
where g=$|g_{ij}|$ is the determinant of metric $g_{ij}$ and \\
$\tilde{\varepsilon}^{\rho\sigma ij}$ =$\left\{\begin{array}{ll}
                                              +1, & \hbox{if $\rho\sigma ij$ is an even permutation of 0123} \\
                                              -1, & \hbox{if $\rho\sigma ij$ is an odd permutation of 0123} \\
                                              0, & \hbox{otherwise.}   \end{array} \right.$
\\Radial component of Magnetic field ($B^{r}$) given by equation (18) can be now rewritten as:
\begin{equation}
B^{r}=\frac{\tilde{\varepsilon}^{tr ij} F_{ij}}{\sqrt{-g}}=\frac{\tilde{\varepsilon}^{tr\theta\phi} F_{\theta\phi}}{\sqrt{-g}}+\frac{\tilde{\varepsilon}^{tr\phi\theta} F_{\phi\theta}}{\sqrt{-g}}
\end{equation}
Using the property of Levi-Civita Symbol ($\tilde{\varepsilon}^{\rho\sigma ij}$) and anti symmetric property of component of electromagnetic field strength tensor $(F_{\theta\phi})$ the above expression can be written as:
\begin{equation}
B^{r}= \frac{2 F_{\theta\phi}}{\sqrt{-g}}
\end{equation}
Thus the difference in Kerr and Kerr-Newman metric lies in replacing $(r_{g} r)$ by $(r_{g} r-Q^{2}-P^{2})$. The determinant of the Kerr-Newman metric can be written as (similar expression is given in case of Kerr metric page 347 (Eqn. 104.5) of Landau and Lifshitz (2008); page 16 (Eqn. 1.70) of Wiltshire et al. (2009)):
\begin{equation}
 g=|g_{i,j}| = -(r^{2}+ a^{2}cos^{2}\theta)^{2} sin^{2}\theta
\end{equation}
Using equations (10-14) we can express the component of electromagnetic field strength tensor $(F_{\phi\theta})$ as:
\begin{equation}
F_{\phi\theta}=\frac{\partial A_{\theta}}{\partial{\phi}}-\frac{\partial A_{\phi}}{\partial{\theta}}=-\frac{\partial A_{\phi}}{\partial{\theta}}
\end{equation}
After simplification the above equation can be written as:
\begin{equation}
 F_{\phi\theta}=-[\frac{Qar sin2\theta-P(r^{2}+ a^{2})sin\theta }{r^{2}+ a^{2}cos^{2}\theta}+\frac{(Q a r sin^{2}\theta + P (r^{2}+ a^{2}) cos\theta) a^{2} sin2\theta
 }{(r^{2}+ a^{2}cos^{2}\theta)^{2}}]
\end{equation}
Combining the equations (21), (22) and (24), the radial component of magnetic field ($B^{r}$) can be written as:
\begin{equation}
B^{r}=-\frac{2}{(r^{2}+ a^{2}cos^{2}\theta) sin\theta}[\frac{Qar sin2\theta-P(r^{2}+ a^{2})sin\theta }{r^{2}+ a^{2}cos^{2}\theta}+\frac{(Q a r sin^{2}\theta + P (r^{2}+ a^{2}) cos\theta) a^{2} sin2\theta
 }{(r^{2}+ a^{2}cos^{2}\theta)^{2}}]
\end{equation}
 Substituting $\theta =\frac{\pi}{2}$ in equations (17) and (25), we can write the radial components of electric and magnetic field for equatorial plane as:
 \begin{equation}
 E^{r}({\theta=\frac{\pi}{2}})=\frac{-Q}{r^{2}}
 \end{equation}
 and
 \begin{equation}
B^{r}({\theta=\frac{\pi}{2}})=\frac{2P(r^{2}+a^{2})}{r^{4}}
 \end{equation}
 \section{\label{sec:Gravito-electro-magneto static redshift} Gravitational and Electro-Magneto-Static Redshift from Rotating Body}
Frame dragging is a general relativistic feature of all solutions to the Einstein field equations associated with rotating masses. Due to the influence of gravity, frame dragging or dragging of inertial frame arises in the Kerr metric. The quantity $\frac{d\phi}{cdt}$ is termed as angular velocity of frame dragging as given by Collas and Klein (2004).\\
Considering the ray of light emitted radially outward from the surface of a compact object (from a rotating body with radius \lq R\rq), the general expression of angular velocity of frame dragging ($\frac{d\phi}{cdt}$) in Kerr field was given as (Eqn. (46) of  Dubey and Sen (2014):
 \begin{equation}
\frac{d\phi}{cdt}( \phi, \theta)=\frac{\frac{R sin\phi sin\theta}{sin^{2}\theta} (1-\frac{r_{g}r}{\rho^{2}})+\frac{r_{g}ra}{\rho^{2}}}{-\frac{r_{g}r a}{\rho^{2}} (R sin \phi sin\theta) + (r^{2}+a^{2}+\frac{r_{g}r a^{2}}{ \rho^{2}}sin^{2}\theta)}
\end{equation}
 If we consider a rotating body having electric charge (Q) and magnetic charge (P), then using the above equation we can write the expression for angular velocity of frame dragging in Kerr-Newman Field ($\frac{d\phi}{cdt}(\phi, \theta)_{KN}$) as:
\begin{equation}
\frac{d\phi}{cdt}(\phi,\theta)_{KN}=\frac{\frac{R sin\phi sin\theta}{sin^{2}\theta} (1-\frac{(r_{g}r-Q^{2}-P^{2})}{r^{2}+ a^{2}cos^{2}\theta})+\frac{(r_{g}r-Q^{2}-P^{2})a}{r^{2}+ a^{2}cos^{2}\theta}}{-\frac{(r_{g}r-Q^{2}-P^{2}) a}{r^{2}+ a^{2}cos^{2}\theta} (R sin \phi sin\theta) + (r^{2}+a^{2}+\frac{(r_{g}r-Q^{2}-P^{2}) a^{2}}{ r^{2}+ a^{2}cos^{2}\theta}sin^{2}\theta)}
\end{equation}
Again for any general $\theta$ and at $\phi=\frac{\pi}{2}$, the expression for $\frac{d\phi}{cdt}$ from above equation (29) can be written as:
\begin{equation}
\frac{d\phi}{cdt}(\phi=\frac{\pi}{2},\theta)_{KN}=\frac{\frac{R}{sin\theta} (1-\frac{(r_{g}r-Q^{2}-P^{2})}{r^{2}+ a^{2}cos^{2}\theta})+\frac{(r_{g}r-Q^{2}-P^{2})a}{r^{2}+ a^{2}cos^{2}\theta}}{-\frac{(r_{g}r-Q^{2}-P^{2}) a}{r^{2}+ a^{2}cos^{2}\theta} (R sin\theta) + (r^{2}+a^{2}+\frac{(r_{g}r-Q^{2}-P^{2}) a^{2}}{ r^{2}+ a^{2}cos^{2}\theta}sin^{2}\theta)}
\end{equation}
For equatorial plane where $\theta =\frac{\pi}{2} $, the above expression (30) can be rewritten as:
\begin{equation}
\frac{d\phi}{cdt}(\phi=\frac{\pi}{2},\theta =\frac{\pi}{2})_{KN} = \frac{R (1-\frac{(r_{g}r-Q^{2}-P^{2})}{r^{2}})+\frac{(r_{g}r-Q^{2}-P^{2})a}{r^{2}}}{-\frac{(r_{g}r-Q^{2}-P^{2}) a}{r^{2}} (R) + (r^{2}+a^{2}+\frac{(r_{g}r-Q^{2}-P^{2}) a^{2}}{ r^{2}})}
\end{equation}
The above expression (31) is the expression for frame dragging $(\frac{d\phi}{cdt})$ on the equatorial plane.\\\\
In General relativity, redshift (Z) and redshift factor ($\Re$) are defined as:
\begin{equation}
\frac{1}{Z+1}= \Re =\frac{\omega}{\omega^{'}}
\end{equation}
 This $\omega^{'}$ is the frequency measured by a distant observer in terms of proper time ($\tau$) and $\omega$ is the frequency measured in terms of the world time (t). A redshift (Z) of zero corresponds to an un-shifted line, whereas $Z<0$ indicates blue-shifted emission and $Z>0$ red-shifted emission. A redshift factor ($\Re$) of unity corresponds to an un-shifted line, whereas $\Re<1$ indicates red-shifted emission and $\Re>1$ blue-shifted emission.\\\\
For a sphere, the photon is emitted at a location on its surface where $dr = d\theta =0$, as the sphere rotates.\\\\
As a result, we can write the expression of frequency as observed by distant observer as:
\begin{equation}
\omega^{'}= -\frac{\partial\Psi}{\partial x^{0}}\frac{\partial x^{0}}{\partial \tau} - \frac{\partial\Psi}{\partial \phi}\frac{\partial \phi}{\partial \tau}=\frac{\omega}{\sqrt{g_{tt}+ g_{\phi\phi}(\frac{d\phi}{c dt})^{2} +2 g_{t \phi}(\frac{d\phi}{c dt})}}
\end{equation}
where $\Psi\equiv -k_{i}x^{i} +\alpha$, is defined as eikonal and $k_{i}$ is the wave four - vector.\\\\
 Following (Eqn.(69) of Dubey and Sen (2014)) we can write the expression for Gravitational Redshift factor for a rotating body with electric charge (Q) and magnetic charge (P) as:
$$\Re(\phi,\theta)_{KN}=$$
\begin{equation}
 \sqrt{{(1-\frac{r_{g} r-Q^{2}-P^{2}}{r^{2}+ a^{2}cos^{2}\theta})-[r^{2}+a^{2}+\frac{(r_{g}r-Q^{2}-P^{2}) a^{2}sin^{2}\theta}{r^{2}+ a^{2}cos^{2}\theta}]sin^{2}\theta}{(\frac{d\phi}{c dt}})^{2} +2 \frac{a sin^{2}\theta (r_{g}r-Q^{2}-P^{2})}{r^{2}+ a^{2}cos^{2}\theta}(\frac{d\phi}{c dt})}
\end{equation}
where $\frac{d\phi}{c dt} \equiv \frac{d\phi}{cdt}(\phi,\theta)_{KN} $ is the corresponding expression of angular velocity of frame dragging in Kerr-Newman Field as given by equation (29).\\
Again for any general $\theta$ and at $\phi=\frac{\pi}{2}$, the expression for $\Re(\phi=\frac{\pi}{2},\theta)_{KN}$ from above equation (34) can be written as:
$$\Re(\phi=\frac{\pi}{2},\theta)_{KN}=$$
\begin{equation}
 \sqrt{{(1-\frac{r_{g} r-Q^{2}-P^{2}}{r^{2}+ a^{2}cos^{2}\theta})-[r^{2}+a^{2}+\frac{(r_{g}r-Q^{2}-P^{2}) a^{2}sin^{2}\theta}{r^{2}+ a^{2}cos^{2}\theta}]sin^{2}\theta}{(\frac{d\phi}{c dt}})^{2} +2 \frac{a sin^{2}\theta (r_{g}r-Q^{2}-P^{2})}{r^{2}+ a^{2}cos^{2}\theta}(\frac{d\phi}{c dt})}
\end{equation}
where $\frac{d\phi}{c dt} \equiv \frac{d\phi}{cdt}(\phi=\frac{\pi}{2},\theta)_{KN} $ is the corresponding expression of angular velocity of frame dragging in Kerr-Newman Field as given by equation (30).\\\\
For equatorial plane where $\theta =\frac{\pi}{2} $, the above expression (35) of $\Re(\phi=\frac{\pi}{2},\theta)_{KN}$ can be written as:
$$\Re(\phi=\frac{\pi}{2},\theta =\frac{\pi}{2})_{KN}=$$
\begin{equation}
 \sqrt{{(1-\frac{r_{g} r-Q^{2}-P^{2}}{r^{2}})-(r^{2}+a^{2}+\frac{(r_{g}r-Q^{2}-P^{2}) a^{2}}{r^{2}})}{(\frac{d\phi}{c dt}})^{2} +2 \frac{a  (r_{g}r-Q^{2}-P^{2})}{r^{2}}(\frac{d\phi}{c dt})}
\end{equation}
where $\frac{d\phi}{c dt} \equiv \frac{d\phi}{cdt}(\phi=\frac{\pi}{2},\theta=\frac{\pi}{2})_{KN}$ is the corresponding expression of angular velocity of frame dragging on the  equatorial plane as given by equation (31).\\\\
If we set magnetic charge (P) and rotation parameter of source (a) both equal to zero in the above expression of redshift factor (36), then the coresponding redshift factor in  Reissner - Nordstrom Geometry ($\Re_{RN}$) can be written as:
\begin{equation}
 \Re_{RN}=\sqrt{1-\frac{r_{g}}{r}+ \frac{Q^{2}}{r^{2}}}=\frac{1}{Z_{RN}+1}
\end{equation}
where $Z_{RN}$ is the gravitational redshift in Reissner - Nordstrom Geometry.\\\\
Now the obtained expression of redshift factor (37) exactly matches with the gravitational redshift factor for a static spherically symmetric electrically charged object in Reissner - Nordstrom Geometry (Eqn. (30) of  Payandeh and Fathi (2013)).\\\\
 We can replace the values of electric charge (Q) and magnetic charge (P) in terms of radial components of Electric field $(E^{r})$ and Magnetic field $(B^{r})$ by utilizing equations (17) and (25). Thus we can obtain the general expression for the Redshift factor from charged rotating body in Kerr-Newman Field, which can be considered as combined Gravitational and Electro-Magneto-Static Redshift factor.\\\\
 In the above expression of redshift factor given by equations (34) to (36):
\begin{itemize}
  \item If we substitute P=0 and a=0 then we can obtain the coresponding redshift factor in  Reissner - Nordstrom Geometry.
  \item If we substitute Q=P=0, then we can obtain the coresponding redshift factor in Kerr Geometry.
 \item If we substitute Q=P=0 and a=0, then we can obtain the coresponding redshift factor in Schwarzschild Geometry.
 \item If we substitute a=0, then we can obtain the coresponding redshift factor from a static body of same mass (Schwarzschild Mass) having static electric and magnetic charge present in the body.
\end{itemize}
Further for showing the physical significance of the calculations reported here, we have considered a pulsar PSR J 1748-2446ad (Dubey and Sen (2014); Hessels et al. (2006)) having Schwarzschild radius ($r_{g}$)=4.05 km, physical radius (R)= 20.10 km, and rotation parameter (a)=2.42 km.\\\\
In Fig. 1, we make a plot using equation (35), showing the variation of redshift ($ Z(\phi=\frac{\pi}{2},\theta)$) with latitude ($\theta$), at different values of $(Q^{2}+P^{2})$ = 0, 1.0 $\times 10^{6}$, 1.0 $\times 10^{7}$, 1.5 $\times 10^{7}$, 1.8 $\times 10^{7}$ $km^{2}$. There are some of arbitrary values of $(Q^{2}+P^{2})$, permissible by equation (9), such that $\triangle> 0$.\\\\
From Fig. 1, it is clearly seen that the value of gravitational redshift increases as the sum of square of electrostatic and magnetostatic charges ($Q^{2}+P^{2}$) increases. The amount of gravitational redshift also increases from pole to equatorial region (maximum at equator) at a fixed value of $(Q^{2}+P^{2})$.\\\\
\begin{figure*}
\includegraphics[width=45pc, height=35pc,angle=270]{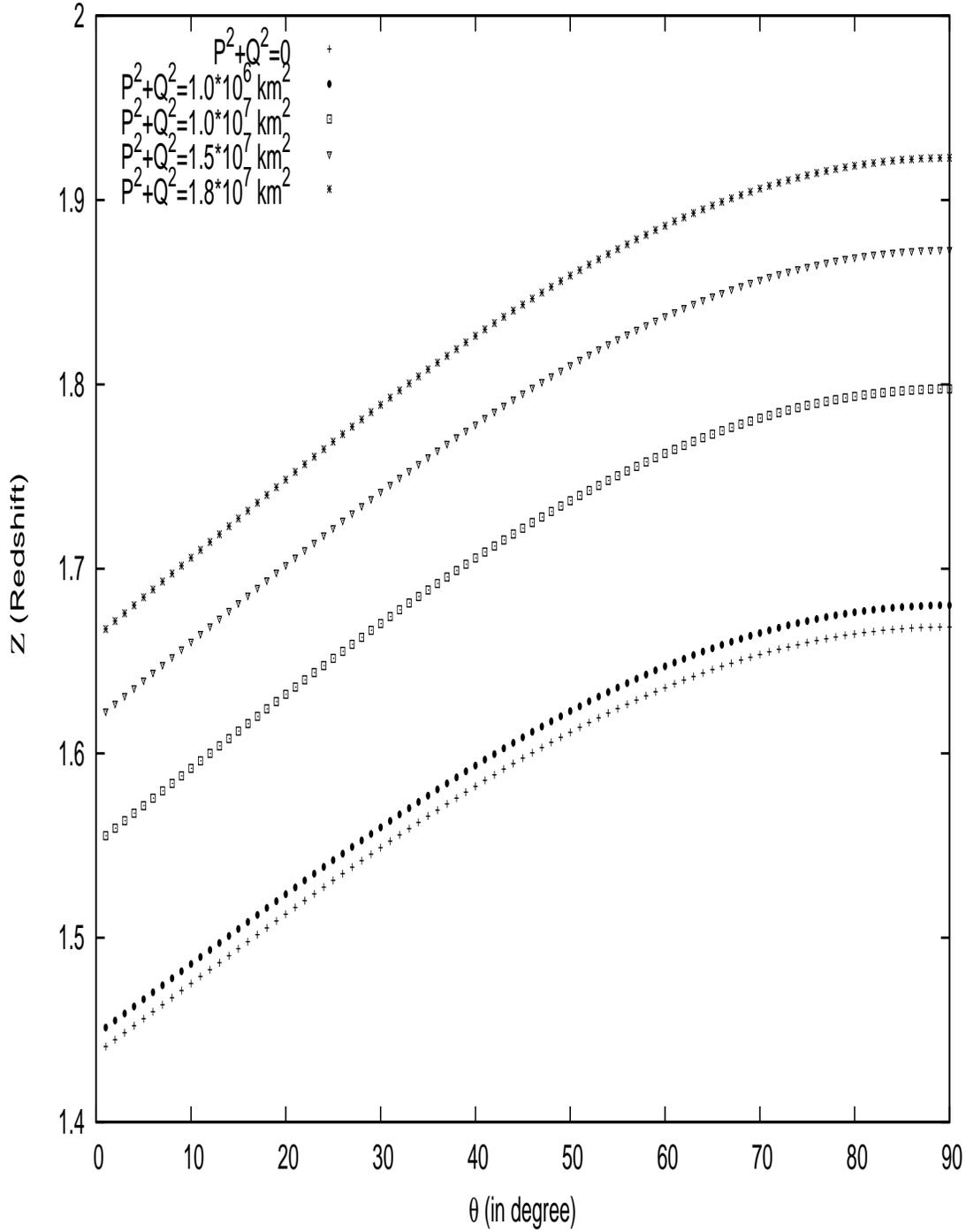}
 \caption{\label{fig1} Shows the variation of redshift
($ Z(\phi=\frac{\pi}{2},\theta)$) versus latitude ($\theta$) from $\theta$ = 0 to $90^o$, at different values of $(Q^{2}+P^{2})$ = 0, 1.0 $\times 10^{6}$, 1.0 $\times 10^{7}$, 1.5 $\times 10^{7}$, 1.8 $\times 10^{7}$ $km^{2}$, for a pulsar PSR J 1748-2446ad.}
\end{figure*}
\begin{figure*}
\includegraphics[width=45pc, height=35pc,angle=270]{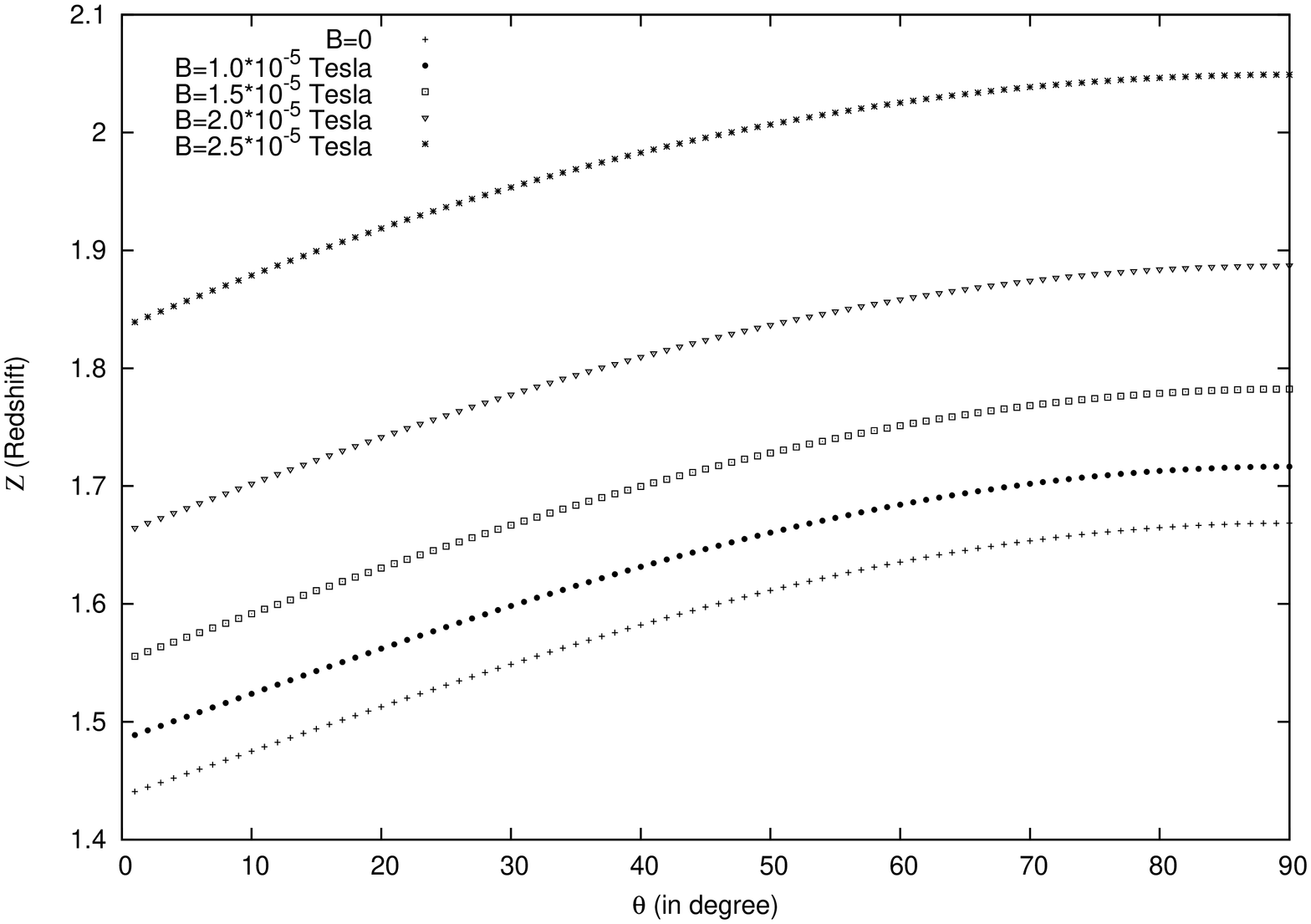}
 \caption{\label{fig1} Shows the variation of redshift ($ Z(\phi=\frac{\pi}{2},\theta)$) versus latitude ($\theta$) from $\theta$ = 0 to $90^o$, at different values of magnetostatic field $(B^{r})$ = 0, 1.0 $\times 10^{-5}$, 1.5 $\times 10^{-5}$, 2.0 $\times 10^{-5}$, 2.5 $\times 10^{-5}$  Tesla, for a pulsar PSR J 1748-2446ad.}
\end{figure*}
 Below we apply the results of our calculations to one practical case.\\\\
 Pulsars are rapidly rotating neutron stars and it is known that neutron stars have intense magnetic field. If we consider a rotating object having intense magnetic field (such as pulsars), and electrostatic charge (Q) equals to zero, then we proceed as follows: \\
  From equation (25), we can rewrite the radial component of magnetic field ($B^{r}(Q=0)$) as:
\begin{equation}
B^{r}(Q=0)=-\frac{2}{(r^{2}+ a^{2}cos^{2}\theta) sin\theta}[\frac{-P(r^{2}+ a^{2})sin\theta }{r^{2}+ a^{2}cos^{2}\theta}+\frac{P (r^{2}+ a^{2}) cos\theta a^{2} sin2\theta
 }{(r^{2}+ a^{2}cos^{2}\theta)^{2}}]
\end{equation}
After simplification the above equation (38) can be rewritten as:
\begin{equation}
B^{r}(Q=0)=\frac{2 P(r^{2}+ a^{2})(r^{2}- a^{2}cos^{2}\theta)}{(r^{2}+ a^{2}cos^{2}\theta)^{3}}
\end{equation}
Now we can write magneto static charge (P) in terms of radial component of magnetic field ($B^{r}$) as:
\begin{equation}
P(\theta)=\frac{B^{r}(r^{2}+ a^{2}cos^{2}\theta)^{3}}{2(r^{2}+ a^{2})(r^{2}- a^{2}cos^{2}\theta)}
\end{equation}
Substituting electrostatic charge (Q) equals to zero in the expression of redshift factor as given by equation (30), we can write the expression of redshift factor ($\Re_{Magnetostatic}(\phi=\frac{\pi}{2},\theta)$) of a rotating object having intense magnetic field as:
$$\Re_{Magnetostatic}(\phi=\frac{\pi}{2},\theta) \ or\ (Z+1)^{-1}=$$
\begin{equation}
 \sqrt{{(1-\frac{r_{g} r-P^{2}}{r^{2}+ a^{2}cos^{2}\theta})-[r^{2}+a^{2}+\frac{(r_{g}r-P^{2}) a^{2}sin^{2}\theta}{r^{2}+ a^{2}cos^{2}\theta}]sin^{2}\theta}{(\frac{d\phi}{c dt}})^{2} +2 \frac{a sin^{2}\theta (r_{g}r-P^{2})}{r^{2}+ a^{2}cos^{2}\theta}(\frac{d\phi}{c dt})}
\end{equation}
where $\frac{d\phi}{c dt} \equiv \frac{d\phi}{cdt}(\phi=\frac{\pi}{2},\theta)_{KN}$ can be obtained by substituting electrostatic charge (Q) equals zero in equation (30).
\begin{equation}
\frac{d\phi}{cdt}(\phi=\frac{\pi}{2},\theta)=\frac{\frac{R}{sin\theta} (1-\frac{(r_{g}r-P^{2})}{r^{2}+ a^{2}cos^{2}\theta})+\frac{(r_{g}r-P^{2})a}{r^{2}+ a^{2}cos^{2}\theta}}{-\frac{(r_{g}r-P^{2}) a}{r^{2}+ a^{2}cos^{2}\theta} (R sin\theta) + (r^{2}+a^{2}+\frac{(r_{g}r-P^{2}) a^{2}}{ r^{2}+ a^{2}cos^{2}\theta}sin^{2}\theta)}
\end{equation}
The obtained expression (41) of redshift factor of a rotating object having intense magnetic field, also shows the dependence on latitude at which light ray has been emitted.\\\\
In Fig. 2, we make a plot using equation (41), showing the variation of redshift ($ Z(\phi=\frac{\pi}{2},\theta)$) with latitude ($\theta$), at different values of magnetostatic field $(B^{r})$ = 0, 1.0 $\times 10^{-5}$, 1.5 $\times 10^{-5}$, 2.0 $\times 10^{-5}$, 2.5 $\times 10^{-5}$  Tesla, for a pulsar PSR J 1748-2446ad. From Fig. 2, it is clearly seen that the value of gravitational redshift increases as the magnetostatic field $(B^{r})$ increases. The amount of gravitational redshift also increases from pole to equatorial region (maximum at equator) at a fixed value of magnetostatic field $(B^{r})$.\\\\
This work has high significance in the area of astrophysics research. There are many objects in nature, like neutron stars, magnetars etc which have high amount of rotation, electric and magnetic field. In general, Sun - like stars have surface magnetic field. In addition one can expect a Sun - like star to hold some amount of net electric charge ($1.5\times10^{28}$ e.s.u) due to frequent escape of electrons than that of protons (Motz (1961)). Thus calculations reported in this paper will have astrophysical significance.
\section{\label{sec:Concl}Conclusions}
\begin{enumerate}
\item Considering the three parameters: mass, rotation parameter and charge, the combined gravitational and
electro-magneto-static redshift factor for a rotating body has been calculated by using Kerr - Newman Geometry.
\item The effect of electrostatic and magnetostatic charges on the amount of redshift of a light ray emitted at various latitudes from the rotating body has been calculated.
\item Under the boundary condition of zero electrostatic and magnetostatic charges, the calculated expression for \lq Gravitational Redshift Factor\rq, reduces to the corresponding expression for gravitational redshift factor in Kerr Geometry. Further if we consider the rotation velocity of the body to be zero, then we can obtain the corresponding gravitational redshift factor for a static body of same mass (Schwarzschild Mass).
  \item  Under the boundary condition of zero magnetostatic charge and zero rotation velocity, the calculated expression for \lq Gravitational Redshift Factor\rq, reduces to the corresponding expression for gravitational redshift factor in Reissner - Nordstrom Geometry. Further if we consider the electrostatic charge to be zero, then we can obtain the corresponding gravitational redshift factor for a static body of same mass (Schwarzschild Mass).
    \item  Gravitational redshift increases as the electrostatic and magnetostatic charges increase, for a fixed value of latitude at which light ray has been emitted.
    \item  Gravitational redshift increases from pole to equatorial region (maximum at equator), for a given set of values for electrostatic and magnetostatic charge.
\end{enumerate}
\textbf{Acknowledgments}
We wish to thank Dr. Atri Deshmukhya, Department of Physics, Assam University, Silchar, India for inspiring discussions. Finally we are thankful to the anonymous referee of this paper, for very useful comments.

\begin{center}
    \textbf{REFERENCES}
\end{center}
Apsel, D.: International Journal of Theoretical Physics \textbf{17}(8), 643 (1978)\\\\
Apsel, D.: General Relativity and Gravitation \textbf{10}(4), 297 (1979)\\\\
Carroll, S.M.: Spacetime and Geometry. An Introduction to General Relativity vol. 1. Pearson,
Addison Wesley, p. 24, 83, 254, 261-262 (2004)\\\\
Collas, P., Klein, D.: General Relativity and Gravitation \textbf{36}(5), 1197 (2004)\\\\
Dubey, A.K., Sen, A.K.: International Journal of Theoretical Physics \textbf{54}(7), 2398 (2014)\\\\
Hessels, J.W., Ransom, S.M., Stairs, I.H., Freire, P.C., Kaspi, V.M., Camilo, F.: Science \textbf{311}(5769),
1901 (2006)\\\\
Kerr, R.P.: Phys. Rev. Lett. \textbf{11}, 237 (1963)\\\\
Landau, L.D., Lifshitz, E.M.: The Classical Theory of Fields vol. 2. 4th revised edn. Butterworth
- Heine- mann, Indian Reprint, p. 65, 357 (2008)\\\\
Motz, L.: Nature \textbf{189}, 994 (1961)\\\\
Newman, E.T., Couch, E., Chinnapared, K., Exton, A., Prakash, A., Torrence, R.: Journal of
mathematical physics \textbf{6}(6), 918 (1965)\\\\
Payandeh, F., Fathi, M.: International Journal of Theoretical Physics \textbf{52}(9), 3313 (2013)\\\\
Wiltshire, D.L., Visser, M., Scott, S.M.: The Kerr Spacetime: Rotating Black Holes in General
Relativity. Cambridge University Press, p. 8 (2009)
\end{document}